# Quantum Computation for Predicting Electron and Phonon Properties of Solids

Kamal Choudhary[1,2]

1. Materials Science and Engineering Division, National Institute of Standards and Technology, Gaithersburg, MD, 20899, USA.

2 Theiss Research, La Jolla, CA, 92037, USA.

**Abstract**

Quantum chemistry is one of the most promising near-term applications of quantum computers. Quantum algorithms such as variational quantum eigen solver (VQE) and variational quantum deflation (VQD) algorithms have been mainly applied for molecular systems and there is a need to implement such methods for periodic solids. Using Wannier tight-binding Hamiltonian (WTBH) approaches, we demonstrate the application of VQE and VQD to accurately predict both electronic and phonon bandstructure properties of several elemental as well as multi-component solid-state materials. We apply VQE-VQD calculations for 307 spin-orbit coupling based electronic WTBHs and 933 finite-difference based phonon WTBHs. Also, we discuss a workflow for using VQD with lattice Green's function that can be used for solving dynamical mean-field theory problems. The WTBH model solvers can be used for testing other quantum algorithms and models also.

1. **Introduction**

Quantum chemistry is one of the most attractive applications for quantum computations[1]. Quantum computers with a few qubits can potentially exceed classical computers because the size of Hilbert space exponentially increases with the number of orbitals[2,3]. Predicting the energy levels of a Hamiltonian is a key problem in quantum chemistry. In the last decade, there has been a significant effort in estimating energies of Hamiltonians[4,5] using various quantum algorithms such as quantum phase estimation (QPE)[1], variational quantum eigen solver (VQE)[6], variational quantum deflation eigen solver (VQD)[7], quantum subspace expansion (QSE),[8] quantum equation of motion (QEOM)[9], quantum amplitude estimation (QAE)[10], witness-assisted variational eigenspectra solver (WAVES)[11], quantum approximate optimization algorithm (QAOA)[12] and quantum annealing (QA)[13]. These algorithms have been used for predicting both ground state and excited states of the Hamiltonians. However, their applications are mainly limited to molecular systems such as $H_2$, LiH, $BeH_2$[6,14-16] molecules and there is a lack of research to apply and evaluate these algorithms for solid-state materials. While the molecular systems have promising applications in the drug-discovery process, the solid-state simulations can accelerate design of superconductors, low-dimensional and topological materials which can improve Noisy Intermediate-Scale Quantum (NISQ) technology[17-19].

Variational quantum eigen solver (VQE) is one of the most celebrated methods for predicting an approximate ground state of a Hamiltonian on a quantum computer following the variational principles of quantum mechanics. VQE is developed as an alternate algorithm of quantum phase estimation (QPE) which solves an eigenvalue of a matrix from a state vector. VQE utilizes Ritz variational principle where a quantum computer is used to prepare a wave function ansatz of the system and estimate the expectation value of its electronic Hamiltonian while a classical optimizer is used to adjust the quantum circuit parameters in order to find the ground state energy. Hence, VQE is a hybrid classical-quantum algorithm because utilizes both the classical and quantum hardware and can circumvent the limited coherence time of currently available quantum circuits. A typical VQE task is carried out as follows: an ansatz with tunable parameters is constructed and a quantum circuit capable of representing this ansatz is designed. The tunable parameters are variationally adjusted until they minimize the expectation value of the Hamiltonian to arbitrary accuracy. Classical computers are used to setup the Hamiltonian terms (such as decomposing Hamiltonian into Pauli matrices) and update the tunable parameters during circuit optimization. Quantum computers are used to prepare a quantum state based on a set of ansatz parameters and perform measurements of interaction terms. More details about VQE can be found in Ref.[20] Significant research work has been done to expand VQE algorithm to evaluated energy levels of Hamiltonian beyond ground state[7,8,21,22]. One of the common methods to estimate high energy levels is Variation Quantum Deflation (VQD), which shows that overlap estimation can be used to deflate eigenstates once they are found, enabling the calculation of excited state energies and their degeneracies. VQD requires the same number of qubits as VQE and is robust for controlling errors[7]. VQD has been successfully used to obtain higher energy levels for several molecular systems[7,23]

In this work, we show that eigen energies of Hamiltonians for solid state materials can be accurately estimated using Wannier tight-binding Hamiltonian approach and variational quantum deflation eigen solver. WFs are a complete orthonormalized basis set that acts as a bridge between a delocalized plane wave representation commonly used in electronic structure calculations and a localized atomic orbital basis that more naturally describes chemical bonds[24-26]. Wannier tight-binding Hamiltonians (WTBH) are a computationally efficient way to calculate properties of materials. One of the most common ways of obtaining Wannier tight-binding Hamiltonians (WTBH)[27-29] is by using density functional theory (DFT) calculations. All major DFT codes support generation WTBHs for a material. WTBHs have been proven successful for accurately predicting both electron, phonon, and electron-phonon coupling and dynamical properties of solids. One of the most popular basis sets for simulating solids is plane-wave basis, but the Hamiltonians using such basis are quite large to simulate on current quantum hardware. Instead, Wannier based approaches provide smaller basis sizes (tens to hundreds of orbitals) which can be simulated on current quantum hardware with tens to hundreds of qubits. Additionally, WTBHs provide dense interpolation on Brillouin zone which is difficult for plane waves even on classical computers. We use the WTBHs for electrons databases[30] for 3D and 2D materials, which was recently populated for thousands of materials using density functional theory approaches. WTBHs can also be used to obtain many other solid-state properties such as optical conductivity, Berry phase, and Chern number[31]. The WTBHs for phonons[32] can be obtained with finite-difference (FD)[33-35] and density functional perturbation theories (DFPT)[36-39]. Such databases have also been populated for thousands of materials, especially for accurate gamma-point phonons[33,36]. Both electron and phonon WTBHs are publicly available in the Joint Automated Repository for Various Integrated Simulations (JARVIS) infrastructure (https://jarvis.nist.gov/) of databases and tools[40]. We integrate these databases with VQD algorithms and simulate thousands of materials using quantum algorithms using JARVIS-tools and Qiskit software[41,42]. In addition to providing an interface to Qiskit, we provide integration with other software such as Tequila[43], and Pennylane[44].

First, we show a detailed analysis of example material, Aluminum and then extend the workflow for 1240 WTBHs for statistical analysis. Although this work deals with single-particle picture, we believe our work can pave the way for solving interacting Hamiltonians (such as dynamical mean-field theory (DMFT)[45] and Green's function and screened Coulomb (GW)[46]), which can be more suitable to simulate on quantum computers than classical computers. We provide a preliminary workflow that can be used to integrate the VQD algorithms with DMFT based solving of lattice Green's function. All the databases and tools from this work are made publicly available so that researchers can apply their own algorithms as well as reproduce our work independently.

2. Methods

Density functional theory calculations for generating electronic and phonon WTBHs were carried out using the Vienna Ab-initio simulation package (VASP)[47,48] software using the workflow given on our JARVIS-Tools[40] GitHub page (https://github.com/usnistgov/jarvis). We use the OptB88vdW functional[49], which gives accurate lattice parameters for both vdW and non-vdW (3D-bulk) solids[50]. We optimize the crystal-structures of the bulk and monolayer phases using VASP with OptB88vdW. The crystal structure was optimized until the forces on the ions were less than 0.01 eV/Å and energy less than $10^{-6}$ eV.

The basic formalism of Wannierization is well-established[31]. Wannier functions for a cell **R** and band n for a given Bloch state $\psi_{n\mathbf{k}}$ and unit-cell volume V is given as:

$$|\mathbf{R}n\rangle = \frac{V}{(2\pi)^3} \int d\mathbf{k} e^{-i\mathbf{k}.\mathbf{R}} |\psi_{n\mathbf{k}}\rangle |\psi_{n\mathbf{k}}\rangle \tag{1}$$

The tight-binding Hamiltonians of electrons and phonons are built with different basis. For electrons, basis orbitals are selected according to the constituent elements, type of pseudopotential chosen and the number of basis orbitals (n per atom). For example, four orbitals (s, $p_x$, $p_y$, $p_z$) are selected as the basis for Aluminum in a system. A full list of orbitals used for different elements can be found in Ref. [30]. The site-site coupling is then described by a n × n matrix and the Hamiltonian $H_k$ by a nN × nN matrix (N atoms per unit cell) for k-point k. For phonons in a 3-dimensional space, the interatomic coupling is described by a 3 ×3 matrix and $H_k$ has a fixed dimension of 3N ×3N. From the symmetry point of view, the basis orbitals are $p_x, p_y, p_z$[32]. Spin-orbit coupling was included in generating WTBHs for electrons but not for phonons. We use Wannier90[51] to construct Maximally-Localized Wannier Functions (MLWF) based TB-Hamiltonians for electrons. For the case of interest in this work, where we wish to describe both the valence and conduction bands near the Fermi level, it is necessary to first select a set of bands to Wannierize, which includes separating the conduction bands from the free-electron-like bands that generally overlap with them in energy[62]. We use the maximum energy difference between DFT and Wannier bands criteria to examine the accuracy of WTBHs obtained from VASP+Wannier90 calculations. For phonons, we use both finite-difference method[52,53] for calculating force-constants of conventional cells which are then processed with Phonopy[54] software. We make efficient python modules (such as `HermitianSolver`) available in JARVIS-Tools so that users can obtain electron or phonon WTBHs for given JARVIS-IDs (if available) and then easily run quantum algorithms on them. The input/output files used in generating the WTBHs are also made publicly available through Figshare API to enhance transparency.

The variational quantum deflation algorithm is based on the combination of variation quantum eigensolver (VQE) and deflation algorithm for finding eigenvalue of a matrix. VQD[7,22] is one of the most straightforward ways to predict eigenstates of a Hamiltonian. In this algorithm, first, we transform the Hermitian Hamiltonian matrix into Pauli operators. The Hamiltonian is now transformed into the following form:

$$H = \sum_{P \in \{I,X,Y,Z\}^{\otimes n}} h_P P \qquad (2)$$

where I, X, Y, Z are single-qubit Pauli operators and $h_P \in \mathbb{R}$ are corresponding coefficients. One of the common ways to achieve this using Qiskit's `qiskit.aqua.operators.weighted_pauli_operator`. After transforming the Hamiltonian, we choose a parametrized quantum circuit $U(\boldsymbol{\theta})$ (shown later) and iteratively optimize the parameter $\boldsymbol{\theta}$ so that energy expectation value for ground state $\langle 0 | U^\dagger(\boldsymbol{\theta}) H U(\boldsymbol{\theta}) | 0 \rangle$ is minimized. Here, $|0\rangle$ denotes the initialized state of the quantum computer. After the minimization we find an optimal parameter $\boldsymbol{\theta}^*$ which gives the approximate eigenstate $|\psi(\boldsymbol{\theta}_0^*)\rangle$. Considering the shape of the input Hamiltonian as $k \times k$ and setting j = 1, we define a new Hamiltonian as:

$$H_j = H + \sum_{i=0}^{j-1} \beta_i |\psi(\boldsymbol{\theta}_0^*)\rangle\langle\psi(\boldsymbol{\theta}_0^*)| \tag{3}$$

Here $\{\beta_i\}$ is set to numbers defined as the difference between eigenvalues $\{\varepsilon_i\}$ and a large number. In this approach, we deflate the Hamiltonian, and perform VQE on the resultant Hamiltonian to find $k$ high energy states.

Parametrized quantum circuits are essential for many quantum algorithms designed for NISQ devices. There are numerous ways to choose a quantum circuit as ansatz for running quantum algorithms. One should avoid superfluous parameters in the circuit resulting in unnecessary gate operations. At the same time, having an insufficient number of independent parameters in the circuit can lead the classical minimization algorithm to fall into false local minima. A careful design of the quantum circuit is therefore essential to make optimal use of current NISQ devices. We choose the ansatz as QISKIT's[41] `EfficientSU2` two-local circuit as the ansatz which the QISKIT documentation proposes as "a heuristic pattern that can be used to prepare trial wave functions for variational quantum algorithms or classification circuit for machine learning." This circuit consists of N + 1 blocks of RY and RZ gates applied to every qubit. These blocks are interlaced with N blocks containing CNOT (q, q') gates for all q < q'. `EfficientSU2` is considered as one of the most expressive and noise-resilient circuits used in several VQE problems[55-57].

## 3. Results and discussion

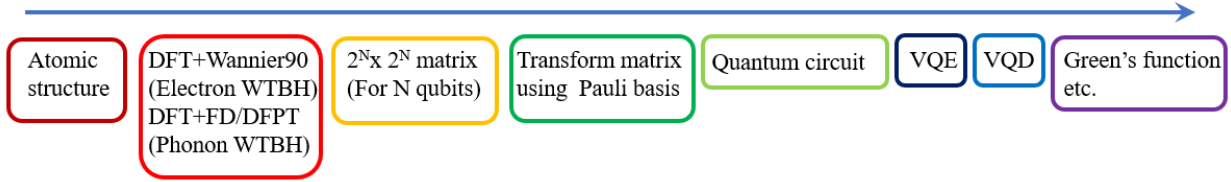

*Fig. 1 Flow chart showing the steps involved in predicting electron and phonon properties of a solid on a quantum computer.*

A typical flowchart for predicting electron and phonon properties using quantum computers is shown in Fig. 1. As a first example of applying a quantum algorithm using Wannier tight-binding Hamiltonian for solids, we compare classical and quantum algorithm-based electronic bandstructure predictions of face-centered cubic Aluminum (JVASP-816). Choosing Al's s, px, py, pz orbitals we obtain an 8x8 Hermitian matrix (i.e. $2^3$ x $2^3$ matrix that can be solved on a 3-qubit quantum computer) at any k-point in the Brillouin zone. This matrix can be easily diagonalized on a classical computer and will act as a reference when comparing with a quantum algorithm. As discussed earlier, the Hermitian matrix is transformed into a sum of Pauli matrices using unitary transformation. The first step in getting the eigenvalues at each k-point is to get the ground state using the variational quantum eigensolver (VQE) algorithm. It is important to note that unlike classical computers, a quantum circuit needs to be made before running quantum algorithms. Quantum circuits can be thought of as the instructions of the quantum system holding all of the quantum operations. Then the coefficients of these matrices are optimized. In a circuit, the qubits are put in order, with qubit zero at the top and qubit two at the bottom and they are read left to right. There are several models in Qiskit for running the quantum algorithms and new models can be easily constructed using available routines. The selection of quantum circuit models are mostly intuitive and depends on the problem of interest[58]. We try 6 ansatzes for Al electronic WTBH. All the 6 circuit models discussed here are made available in the JARVIS-Tools in the `jarvis.core.circuits.QuantumCircuitLibrary` module. Here, circuit-4, circuit-5 and circuit-6 are also known as `RealAmplitudes`, `PauliTwoDesign` and `EfficientSU2` circuits. RY and RZ represent parametrized circuits with parameters ө. In Fig. 3a, we observe all the 6 circuits can accurately predict eigenvalues at Gamma point for Al electronic WTBH. However, only 6 can predict accurate energy levels at X-point. Here, the green lines denote the reference energy level from classical Numpy eigenvalue solvers. We note that circuit 6 is more complex in terms of gates and associated parameters compared to others. However, for general purpose use throughout the dataset, we use circuit-6 as the circuit model. There are multiple input parameters for circuit-6 out of which the number of qubits and number of repeat units are the most critical ones for this work. Although circuit-6 with one repeat unit can be used to simulate face-centered cubic (FCC) Aluminum (JVASP-816) WTBH, we find the number of repeat units need to be increased for more complex systems such as hexagonal PbS (JVASP-35680) as shown in Fig. 3c. We observe that at least 4 repeat units are necessary for getting accurate energy levels for PbS at X point. Hence, for generalizability purposes, from here on we only use circuit-6 model with 5 repeat units. An algorithm to automatically identify the most accurate circuit model with the least number of parameters would make the whole workflow much faster. However, such automatic workflow development task will be taken up in the future and is beyond the scope of this work.

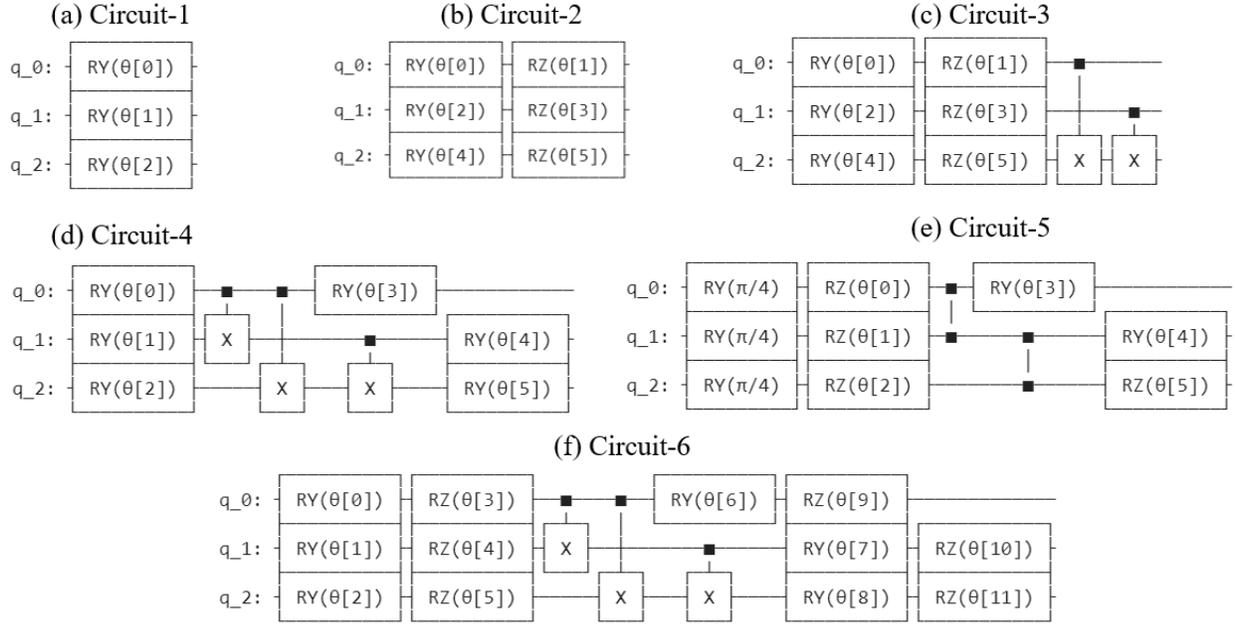

*Fig.2 A set of circuit models considered for simulating example Aluminum electronic WTBH. Here, circuit-4, circuit-5 and circuit-6 are also known as RealAmplitudes, PauliTwoDesign and EfficientSU2 circuits. RY and RZ represent parametrized circuits with parameters θ. All the circuit diagrams are generated using Qiskit. Wires and boxes with 'X' in it represent Controlled-X gate. Wires with two solid squares such as in Circuit-5 represent Controlled-Z gates.*

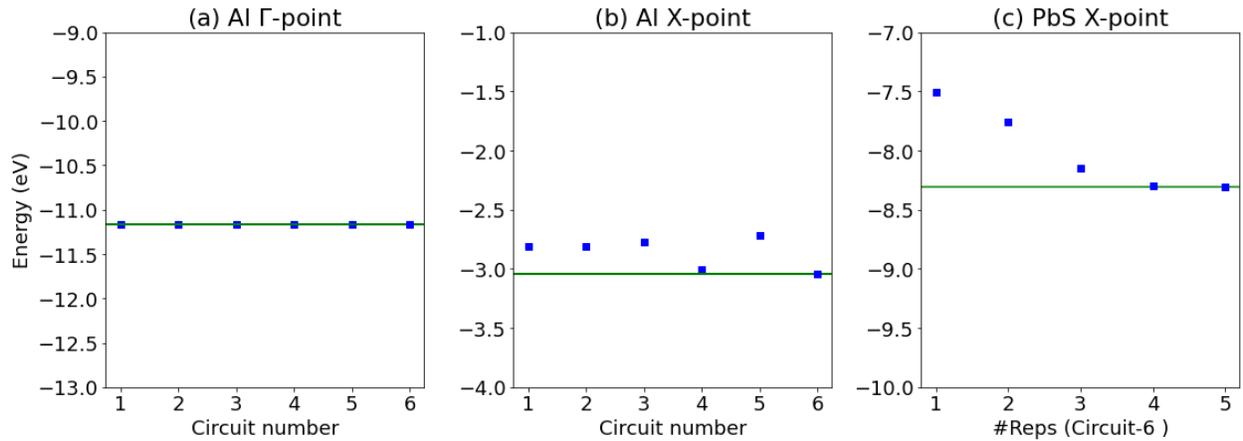

*Fig. 3 Ground state energy predictions for several Kpoints in Brillouin zone for FCC Al and number of circuit-6 repeat unit dependence for hexagonal PbS using electronic WTBHs. a) Al for Gamma point, b) Al for X point, c) PbS for X point for different repeat units of Circuit-6.*

Next, we show an example code snippet to use the WTBH with the circuit-6 model as shown in Fig.4. The WTBH for both phonons and electrons can be easily obtained from the Figshare API with the help of JARVIS-Tools. In the example, we first obtain the electronic WTBH for Al (JARVIS-ID: JVASP-816) at X point in the Brillouin zone and then use circuit-6 with a statevector-simulator backend to get VQE ground state energy level. The backend can be easily replaced with other simulators or real devices to simulate the system. Also, if necessary, the Qiskit circuits can be easily transformed into other quantum circuit packages such as Google's Cirq package.

```python
# pip install -U jarvis-tools qiskit
from jarvis.db.figshare import get_wann_electron, get_wann_phonon, get_hk_tb
from jarvis.io.qiskit.inputs import HermitianSolver
import numpy as np
import qiskit
from qiskit import Aer
from jarvis.core.circuits import QuantumCircuitLibrary
# Aluminum JARVIS-ID: JVASP-816
wtbh,Ef,atoms = get_wann_electron(jid="JVASP-816")
kpt = [0.5, 0., 0.5] # X-point
hk = get_hk_tb(w=wtbh, k=kpt)
HS = HermitianSolver(hk)
n_qubits = HS.n_qubits()
reps = 2
circ = QuantumCircuitLibrary(n_qubits=n_qubits,reps=reps).circuit6()
backend = Aer.get_backend("statevector_simulator")
en, vqe_result, vqe = HS.run_vqe(mode='min_val',var_form=circ,backend=backend)
vals,vecs = HS.run_numpy()
params=vqe.optimal_params
circuit=vqe.construct_circuit(params)
print('Classical, VQE (eV):', (vals[0]-Ef), (en-Ef))
print('Show model\n', circuit[0])
```

*Fig. 4 Example code-snippet from JARVIS-Tools showing how to find lowest eigenvalue for Al at X -point using electronic WTBH and circuit-6.*

In addition to quantum circuit model selection, an appropriate classical optimizer needs to be selected for running VQE algorithm. We optimize the circuit parameters using classical optimizers such as constrained optimization by linear approximation (COBYLA), limited-memory Broyden–Fletcher–Goldfarb–Shanno bound (L_BFGS_B), sequential least squares programming (SLSQP), conjugate gradient (CG), and simultaneous perturbation stochastic approximation (SPSA). These are local optimizers that attempt to find an optimal value within the neighboring set of a candidate solution[59]. COBYLA is an iterative nonlinear derivative–free constrained optimization[60] algorithm that uses a linear approximation approach[61]. The algorithm is easy to use for small numbers of variables. L-BFGS is a quasi-Newton method-based algorithm that approximates the Broyden–Fletcher–Goldfarb–Shanno algorithm (BFGS) using a limited amount of computer memory[62] and is widely used for machine learning tasks. SLSQP optimizer is a sequential least squares programming algorithm which uses the Han–Powell quasi–Newton method with a BFGS update of the B–matrix and an L1–test function in the step–length algorithm[63].CG is a widely used algorithm for solving large-scale unconstrained optimization problems[64]. SPSA is a descent-based method which can also find global minima similar to simulated annealing. It requires only two measurements of the objective function, regardless of the dimension of the optimization problem[65].

We monitor the convergence iterations for several optimizers in Fig. 5a for Al electronic WTBH at X-point. For the particular case, we observe that COBYLA converges fastest and CG the slowest. SLSQP is another suitable optimizer which converges comparably to COBYLA. After obtaining the ground state eigenvalue and state using VQE, we obtain higher energy levels using Variational Quantum Deflation (VQD) algorithm. The same procedure was repeated for all the k-points leading to the electronic and phonon bandstructures for Al as shown in Fig. 5b and 5c. These simulations are carried out on IBM's `statevector_simulator`. We compare the Numpy[66] based eigenvalue solver also to compare with VQE-VQD results. In both of the bandstructure predictions, we observe excellent agreement among them suggesting that WTBH can successfully be used to predict bandstructures with high accuracy.

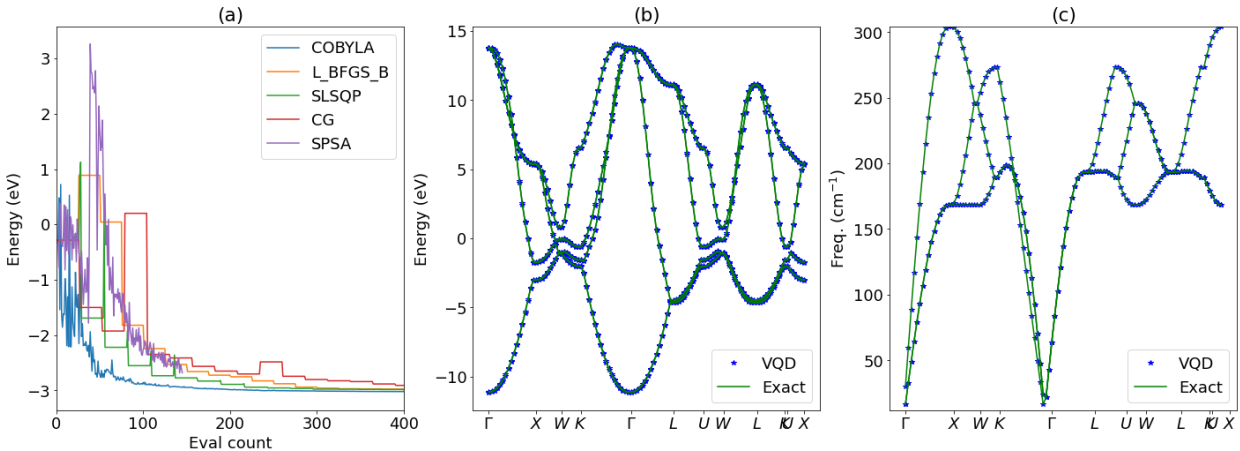

*Fig.5 a) Monitoring VQE optimization progress with several local optimizers such COBYLA, L_BFGS_B, SLSQP, CG, and SPSA for Al electronic WTBH and at X-point. b) electronic bandstructure calculated from classical diagonalization (Numpy-based exact solution) and VQD algorithm for Al. b) phonon bandstructure for Al.*

To further check the validity of the workflow, we screen materials for which the number of orbitals was less than 32 ($2^5$; 5 qubits) and calculate the minimum and maximum allowed energy level with VQE for both phonons and electrons. We compare these values with ground state energy from classical Numpy[66] eigensolvers and the results are shown in Fig. 4. We apply VQE-VQD calculations for 307 spin-orbit coupling based electronic and 933 finite-difference based WTBHs from JARVIS-DFT database. Electronic WTBHs have 16 % unary, 53.36 % binary, 29 % ternary and 2 % quaternary compounds while phonon WTBHs have 0.76 % unary, 31.97 % binary, 66.50 % ternary and 0.76 % ternary compounds suggesting a reasonable vast chemical space. These WTBHs along with the scripts to run the quantum algorithms are made available using JARVIS-Tools. We find the mean absolute error (MAE) for electronic WTBH predictions as 0.024 eV and for phonon 0.1 cm$^{-1}$ with $r^2$=0.999 in both cases suggesting excellent agreement between the quantum algorithms and classical predictions. We provide the all the predictions in the supplementary information (Table S1 and Table S2). While electronic bandstructures are key properties for solids, several other key electronic properties can be predicted using the WTBHs and quantum algorithms such as surface bandstructure, Chern number, Berry curvature, optical conductivity, thermoelectric coefficients[31]. Also, the WTBHs used here are single-particle description of solids but many body picture can be constructed for methods such as Wannier-based dynamical mean-field theory (DMFT)[45], GW[46], and time-dependent Wannier functions[67]. Importantly, we note that the current WTBH for electrons are all based on the single-particle picture and do not predict many-body excited states[68-70].

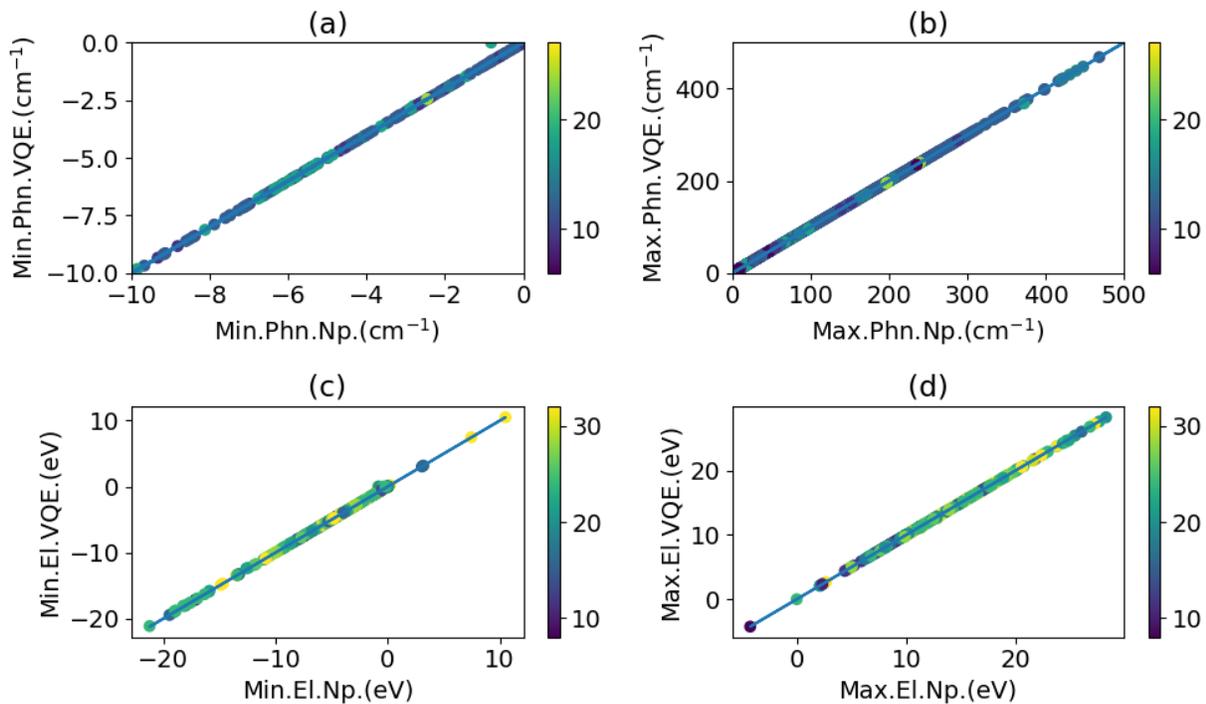

*Fig. 6 Comparison of minimum (Min.) and maximum (Max.) energy levels at Γ-point for electronic and phonon WTBH using classical eigenvalue routine in Numpy (Np.) and VQE solver. a) and b) comparison of phonon (Phn.) minimum and maximum energy levels for 930 materials, c) and d) comparison of electronic (El.) minimum and maximum energy levels for 300 materials. The colorbar represents the number of Wannier orbitals. The data for this plot is given in the supplementary information (Table S1 and Table S2).*

Dynamical mean-field theory (DMFT)[45,71] is one for the commonly used techniques for solving predicting electronic structure of correlated systems using impurity solver models. DMFT maps a many-body lattice problem to a many-body local problem with impurity models. In DMFT one of the central quantities of interest is the Green's function such as:

$$G(k, \omega_n) = [\omega_n + \mu - H(k) - \Sigma(\omega_n)]^{-1} \quad (4)$$

where $\omega_n = (2n + 1)\pi T$ is a fermionic Matsubara frequency at temperature $T$, $\mu$ is the chemical potential, $H(k)$ is a diagonal matrix containing the electronic WTBH eigenvalues (obtained with VQD algorithm mentioned above) at k-point $k$, and $\Sigma$ is non-diagonal part and represents the self-energy. As an initial guess, we set $\Sigma = 0$ and calculate local Green's function for k-points on a dense k-point grid (31x31x31) and frequency grid (-20 eV to 5 eV with 0.25 eV step-size). Now the spectral function ($A$) and DMFT hybridization function ($\Delta$) is calculates as:

$$A(\omega) = -\frac{1}{\pi} \sum_k Im(G(\omega + i\delta)) \quad (5)$$

and

$$\Delta(\omega + i\delta) = \omega - (G)^{-1} \quad (6)$$

with $\delta \to 0$. Many quantum Monty Carlo (QMC) impurity solvers[72,73] take the hybridization function as the input to calculate self-energy. The Green's function is then solved iteratively until the impurity Green's function coincides with that of local lattice Green's function. The final self-energy found after the self-consistency cycle can be used to predict the desired spectral function. An example for the imaginary part of Al's DMFT hybridization function for a few components considering zero self-energy is shown below in Fig. 5.

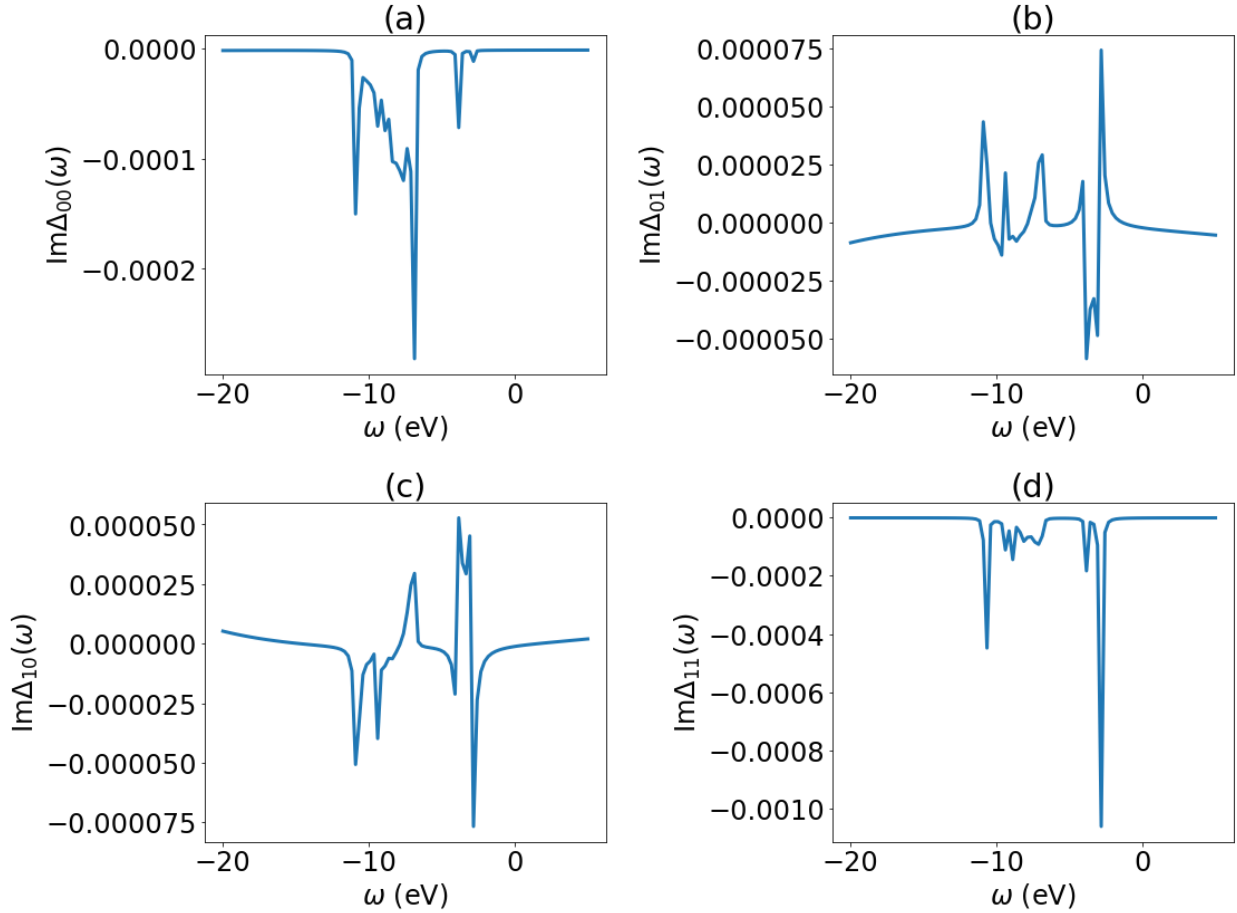

*Fig. 7 Imaginary part of Al's DMFT hybridization function for a few components considering zero self-energy. a)$\Delta_{00}$, b)$\Delta_{01}$, c)$\Delta_{10}$, d)$\Delta_{11}$.*

## 4. Conclusions

In conclusion, we demonstrate the applications of VQE and VQD to accurately predict both electronic and phonon properties of several solid-state materials using Wannier tight-binding

Hamiltonian approach. The WTBH models can act as the testbed for many other quantum algorithms. Although the present work has been used with WTBH, our workflow can be extended for solving Hermitian matrices from plane-wave DFT codes as well, given a larger number of qubits. Also, as Wannier based approaches have already been used to solve time dependent and many-body problem on classical computers, WTBH with quantum algorithms can be useful to tackle problems which are intractable for classical computers.

**Data availability**

The electronic and phonon WTBHs used in this work are available at the JARVIS-DFT website (https://jarvis.nist.gov/jarvisdft/) . Tools to generate quantum circuits using these WTBHs are available at JARVIS-Tools GitHub page (https://github.com/usnistgov/jarvis). The VQE and classical predictions are also provided in the supplementary information.

**Acknowledgements**

K.C. thanks Kevin Garrity from National Institute of Standards and Technology (NIST), Alan Aspuru-Guzik from University of Toronto, and Kristjan Haule from Rutgers University for helpful discussion. The author thanks NIST for computational and funding support.